\documentclass[aps,prb,twocolumn,groupedaddress]{revtex4}  
\usepackage{graphicx}  
\newcommand{\nn}{\nonumber}  
\newcommand{\be}{\begin{equation}}  
\newcommand{\ee}{\end{equation}}  
\newcommand{\bea}{\begin{eqnarray}}  
\newcommand{\eea}{\end{eqnarray}} 
 
\newcommand{\bscco}{BiSr$_2$Ca$_2$CuO$_{8+\delta}$\,}
\newcommand{\hb}{\hbar} 
\newcommand{\la}{\langle} 
\newcommand{\ra}{\rangle} 
\newcommand{\half}{\frac{1}{2}\,}

\begin{document} 
\title{Langevin vortex dynamics for a layered superconductor in the  
lowest Landau level approximation} 
 
\author{W.~A.~Al-Saidi} 
\email{Al-Saidi.1@osu.edu} 
\author{ D.~ Stroud} 
\email{Stroud@mps.ohio-state.edu} 
\affiliation{Department of Physics, 
The Ohio State University, Columbus, Ohio 43210} 
 
\date{\today} 
 
\begin{abstract} 
 
We have numerically investigated the dynamics of vortices in a clean
layered superconductor placed in a perpendicular magnetic field.  We
describe the energetics using a Ginzburg-Landau free energy functional
in the lowest Landau level  approximation.  The dynamics are
determined using the time-dependent Ginzburg-Landau approximation, and
thermal fluctuations are incorporated via a Langevin term.  The c-axis
conductivity at nonzero frequencies, 
as calculated from the Kubo formalism, shows a strong
but not divergent increase as the melting temperature $T_M$ is
approached from above, followed by an apparently 
discontinuous drop at the vortex lattice freezing temperature.  
The discontinuity is consistent
with the occurrence of a first-order freezing.
The calculated equilibrium properties agree
with previous Monte Carlo studies using the same Hamiltonian. 
We briefly discuss the possibility of detecting
this fluctuation conductivity experimentally.
\end{abstract} 

 
\maketitle 
 
\section{Introduction} 
 
Vortices in the mixed state of a clean type-II superconductor are
believed to break the translational symmetry and form a triangular
Abrikosov lattice for magnetic field $B$ exceeding the lower critical
field $H_{c1}$.  At a sufficiently high temperature $T$, thermal
fluctuations are expected to melt this lattice and restore the
translational symmetry through a solid-liquid phase transition.
However, in most low-$T_c$ superconductors, this melting occurs near
the upper critical field $H_{c2}$, and is thus difficult to
distinguish from the usual superconducting-normal transition.  In the
high-T$_c$ cuprates, however, this melting transition is typically
well separated from $H_{c2}$.
 
Many experiments suggest that this solid-liquid phase transition is
{\em first order}.  For example, the resistivity of untwinned single
crystals of YBa$_2$Cu$_3$O$_{7-\delta}$ (YBCO) drops sharply at a
temperature $T_M$ well below the $H_{c2}(T)$ line. \cite{kwok1} This
temperature also coincides with a discontinuous jump in the
magnetization. \cite{kwok2,liang} Most distinctively, both a latent
heat and a specific heat discontinuity have been observed at the
transition.  These signatures have been seen in untwinned single
crystals of YBCO for magnetic fields both parallel and perpendicular
to the c-axis. \cite{kwok3}
 
The first order nature of the transition is supported by a number of
theoretical models. \cite{ sasik1,sasik1a,hu, nordborg, moore} At
least at high fields, this transition is thought to represent a
simultaneous melting of the vortex lattice in the ab-plane and
decoupling of the vortex ``pancakes'' in different layers.  For
example, numerical studies of layered superconductors, using Monte
Carlo methods applied to a model Hamiltonian in the lowest Landau
level (LLL) approximation, suggest a simultaneous melting and
decoupling transition.\cite{sasik1, sasik1a, hu} A similar conclusion
is also suggested by studies of melting using an analogy with a $2$D
Bose system.\cite{nordborg} On the other hand, some workers have
suggested that the LLL actually gives no phase transition at all, but
only a crossover associated with interlayer decoupling. \cite{moore}
  
In this paper, we extend previous numerical studies of flux lattice
melting to treat the {\em dynamics} of a vortex system.  Our
calculations are based on a Lawrence-Doniach model for the free energy
functional of a layered superconductor, treated in the LLL
approximation, and are carried out as a function of temperature at
fixed magnetic field.  The dynamics are treated within the
time-dependent Ginzburg-Landau approximation, with Langevin noise
included to simulate the effects of thermal fluctuations.  A previous
calculation, for a similar model in two dimensions, and with spherical
rather than periodic boundary conditions, has been carried out by
Kienappel and Moore \cite{moore2}.  The LLL approximation is expected
to be most accurate at strong magnetic fields ($H_{c2}/3 < B <
H_{c2}$), but may have a slightly broader range of validity at low
temperatures, since a weak participation of higher Landau levels at
such temperatures can be incorporated by a suitable renormalization of
the LLL model parameters.\cite{tesanovic} The LLL approximation fails,
however, at weak magnetic fields, because it omits the effects of
thermally induced vortex-antivortex pairs.  Consistent with previous
Monte Carlo studies, we find a single first-order liquid-solid phase
transition with simultaneous loss of in-plane and interplane vortex
order.  However, the Langevin simulation also yields information about
dynamical properties such as the conductivity.  In particular, we find
that the c-axis conductivity shows a striking, but not divergent,
increase as the first-order melting temperature $T_M(B)$ is approached
from above.
 
The remainder of this paper is organized as follows.  In Sec. II we
describe the Langevin model, and our method for calculating various
static and dynamic quantities from the model. Our results are
presented in Sec.\ III, followed by a brief discussion in Sec.\ IV.
 
\section{Formalism} 
 
\subsection{Model Hamiltonian and Dynamical Equations} 
 
We consider a three-dimensional ($3$D)  superconductor  
consisting of a  
stack of Josephson- or proximity-coupled $2$D layers.   
We assume that this 
system is described by the free energy functional   
\be 
{\mathcal{F}}= d_0 \sum_{n} \int d^2\,{\bf r} f_n[\psi_n({\bf r})]. 
\label{eq:LG}  
\ee 
Here $n$ is the layer index, $d_0$ is the thickness of one 
layer, and 
\bea  
f_n[\psi_n({\bf r})] &= &\alpha(T) |\psi_n({\bf r})|^2 +\half \beta
|\psi_n({\bf r})|^4 \nn \\
&+&\frac{1}{2\,m_{ab}} \left| \left(-i \hb \nabla - \frac{q\,{\bf 
A}}{c}\right)\psi_n({\bf r})\right|^2 \nn \\ 
&+&
\frac{\hb^2}{2\, m_c d^2} \left| e^{- i \chi_{n,n+1}}\,\psi_{n+1}({\bf r})-\psi_n ({\bf 
r})\right|^2.  
\eea 
$\psi_n({\bf r})$ is the order parameter of the $n$th layer, $q= 
-2e$ is the charge of a Cooper pair, $d$ is the distance between the 
layers, and $\alpha(T)$, $\beta$ , $m_{ab}$, and $m_{c}$ are 
material-dependent parameters. 
The phase factor $\chi_{n,n+1} = 
\frac{2 \pi}{\Phi_0} \int_{n d}^{(n+1) d} dz A_z$, where  
$\Phi_0 = hc/2e$ is 
the flux quantum, and ${\bf A}$ is the vector potential.  We will 
assume that the external magnetic field  
${\bf B} \| z$, i.e. perpendicular to the layers, so that 
$\chi_{n,n+1} = 0$, 
and we choose a gauge such that ${\bf A}= - B y\hat{x}$.    
We also neglect screening currents and fluctuations of 
the vector potential, so that the local and externally applied 
magnetic fields are equal.   
This should be a good approximation when the Ginzburg-Landau  
parameter $\kappa \gg 1$, as in the cuprate superconductors.   
Finally, we assume that ${\bf B}$ is uniform throughout the  
superconductor. 
 
In the LLL approximation, the order   
parameter in each layer is expanded as 
\be 
\psi_n({\bf r})=\psi_0 \sum_{k} c_{n,k} \phi_k(x,y), \label{eq:LLL} 
\ee 
where 
\be 
\phi_k(x,y)= e^{i k x} \, \exp[-(y-k\ell^2)^2/(2\,\ell^2)], 
\ee 
are the lowest eigenstates of the  kinetic energy operator  
$ \left(-i \hb \nabla- {q\,{\bf A}}/{c}\right)^2 /(2m^{*}_{ab})$, 
corresponding to eigenvalue $\hb q B/(2m_{ab}c)$.  
Here 
$\psi_0=\left[\pi \ell^2 |\alpha_H(T)|^2/(\ell_{0}^{2}\, \beta^2) 
\right]^{1/4}$, $\alpha_H(T)=\alpha(T)(1-B/H_{c2})$, $\ell=(|q| 
B/\hb c)^{-1/2}$ is the magnetic length, and  $\ell_0= (4 
\pi/\sqrt{3})^{1/2}\,\ell$.  The magnitude of $\psi_0$ is 
chosen so that the spatial average of $|\psi_n({\bf r})|^2$ 
is $|\alpha_H|/(\beta \beta_\triangle)$ with
$\beta_\triangle=1.169\ldots$  
when the vortices are arranged in a triangular lattice. 
 
We assume the system is a parallelepiped of dimensions $L_x$, $L_y$,
and $L_z$, with periodic boundary conditions in all three directions.
We choose $L_z = N_zd$, where $N_z$ is an integer.  The periodicity condition in the x-direction
$\psi_n(x+L_x,y)=\psi_n(x,y)$ implies $k = 2\pi\,m/L_x$,
where $m$ is an integer.  If each layer contains $N_\phi$ vortices,
there will be $N_\phi$ independent $c_{n,k}$'s labeled by
$m=0,\ldots, (N_\phi-1)$.  The periodicity constraint in the
y-direction, $|\psi(x,y+L_y)|=|\psi(x,y)|$, implies that
$c_{n,m'}=c_{n,m}$ for all $m'=m$ modulo $N_\phi$.  Finally, the
periodicity constraint in the z-direction implies that
$c_{n+N_z,k}=c_{n,k}$.  Besides these periodicity conditions, the cell
dimensions in the x- and y-direction are chosen to be compatible with
a possible triangular lattice.  This choice may be written as
$L_x/L_y= 2 n_x/(\sqrt{3} n_y)$ where $n_x$ and $n_y$ are the number
of vortices along a given row or column parallel to the x- or
y-direction, and $N_\phi = n_xn_y$.
 
Using Eq.\ (\ref{eq:LLL}) for the order parameter,  we can  
rewrite Eq.\ (\ref{eq:LG}) as: 
\be 
{\mathcal{F}}= 
\sum_{n}\left({\mathcal{F}}_{2D}^{(n)}+{\mathcal{F}}_{C}^{(n)} \right), 
\label{eq:ham1} 
\ee 
where ${\mathcal{F}}_{2D}^{(n)}$ is the free energy per layer, and 
${\mathcal{F}}_{C}^{(n)}$ is the coupling between the 
$n$th and $(n+1)$th layers.   These terms take the form 
\bea 
{\mathcal{F}}_{2D}^{(n)}/(k_B T)&=&g^2(B,T)\, n_x \Big[
\mathrm{sgn}(\alpha_H) \sum_{k} |c_{n,k}|^2  \nn \\
&+& \frac{1}{4}\, \sum_{k,p,q} v(p,q)\, c^{}_{n,k} \, c^{*}_{n,k+p}\,c^{*}_{n,k+q} 
\,c^{}_{n,k+p+q}\Big] 
\eea 
and 
\be 
{\mathcal{F}}_{C}^{(n)}/(k_B T)=g^2(B,T)\, n_x \eta \sum_{k} 
|c_{k,n}-c_{k,n+1}|^2.  
\ee 
Here, we have defined 
\be 
v({p,q})= \sqrt{2 \pi \ell^2/\ell_{0}^{2}} 
\exp\left[-\ell^2 \left(p^2+q^2\right)/2\right], 
\ee 
\be 
g^2(B, T)= \pi \ell^2 d_0 \alpha_{H}^{2}/(\beta\,k_B\,T), 
\label{eq:g2} 
\ee 
and introduced the dimensionless interlayer coupling strength 
$\eta=J/|\alpha_H|$ where $J\equiv \hb^2 /( 2 m_c d^2)$ is the
Josephson coupling between the layers. 
The quantity $\mathrm{sgn}({\alpha_H})=-1$ or $+1$ in the 
superconducting or normal regimes; the mean field  
instability occurs when $\alpha_H(T)=0$.   

The parameter $g^2(B, T)$ represents the ratio of the superconducting
condensation energy per vortex per layer $\pi\ell^2 d_0
\alpha_H^2/\beta$ to the thermal energy $k_BT$ within the
Ginzburg-Landau approximation.  Note that, for fixed $\alpha_H$,
$g^2(B, T)$ varies {\em inversely} with temperature.  Thus, a plot of
system properties as a function of $|g|$, may be viewed as a plot as a
function of $T$ for fixed $B$; however, {\em small} $|g|$ represents
{\em large} $T$ (vortex liquid phase).  Previous
calculations\cite{sasik1a,hu} have provided evidence that there is a
first-order vortex lattice melting transition as a function of $g^2$.

We study the dynamics of this system using 
the time dependent Ginzburg-Landau (TDGL) equation in the 
presence of a Langevin noise term.  We write this equation as 
\be 
\Gamma \frac{\partial \psi_n({\bf r},t)}{\partial t}= - \frac{\delta 
{\mathcal{F}}}{\delta{\psi_{n}^{*}({\bf r},t)}} + \xi_n({\bf r},t), 
\label{eq:tdgl} 
\ee 
where $\Gamma$ is the relaxation time parameter, and 
$\xi({\bf r},t) $ is  a white noise term characterized by 
the correlation functions 
\bea 
\la \xi_n({\bf r},t) \ra &=& 0 \nn \\ 
\la \xi_{n}^{*}({\bf r},t) \, \xi_{n'}({\bf r'},0)\ra &=& 2\, k_BT\, \Gamma\, 
\delta({{\bf r -r'}}) \delta_{n,n'}\delta(t), \nn  
\eea 
where $\la \cdots \ra $ denotes an ensemble average. 
We assume that $\Gamma$ is real because the system 
has particle-hole symmetry; with our choice of units, $\Gamma$ 
has the same dimensions as $\hb$.  The noise 
term ensures that the system will remain in a steady state 
at temperature $T$.   
 
Langevin dynamical calculations have previously been carried out by
Ryu and Stroud \cite{ryu} to study vortex lattice melting for both
clean and dirty high-T$_c$ layered superconductors.  They differ from
the present calculation by using a different
equilibrium free energy functional ${\cal F}$ than ours.  
In the
model of Ref.\ \onlinecite{ryu}, the flux lines cannot be broken; this feature should
lead to rather different results from those obtained in the present
model calculations.
 
In the present LLL expansion, the TDGL equation may be rewritten as 
\bea 
\frac{d c_{n,k}}{d\tau}=&-& 
\Bigg[\mathrm{sgn}(\alpha_H)\, c_{n,k}\nn \\ 
&+& \half \, 
\sum_{p,q} v(k-p,q)\, {c_{n,p+q}^{*}} c^{}_{n,k+p} c^{}_{n,q} \Bigg] \nn \\  
 &-& \eta \, (c_{n-1,k}- 2 c_{n,k}+c_{n+1,k})+ {\xi'}_{n,k}(\tau). 
\label{eq:langevin}  
\eea 
Here we have introduced a dimensionless time variable  
$\tau = |\alpha_H|t/\Gamma \equiv t/\,t_0$, 
where $\,t_0= \Gamma/|\alpha_H|$ is a characteristic relaxation 
time.  The noise term $\xi'$ is now described by the correlation 
functions \cite{note0}
\bea 
\la {\xi'}_{n,k}(\tau) \ra &=& 0 ;\\ 
\la {\xi'}_{n,k}^{*}(\tau) \, {\xi'}_{n',k'}(\tau') \ra &=& \frac{2}{n_x\,g^2(B,T)} \delta_{k,k'} \delta_{n,n'}\delta(\tau-\tau'). 
\eea

\subsection{Calculated Quantities} 
 
\subsubsection{Equilibrium Quantities} 
 
Equilibrium quantities can be computed as time averages of the
solutions to the TDGL equations, either in the solid or the liquid
phase.  According to the ergodic hypothesis, this procedure should
give the same results as an equilibrium average obtained by treating
Eq.\ (\ref{eq:ham1}) as a Hamiltonian.  We have, in fact, confirmed
this point by comparing some of our results with those obtained
earlier by other workers from Eq.\ (\ref{eq:ham1}) using Monte Carlo
techniques.\cite{hu, sasik1}
 
We have evaluated several thermodynamic properties 
of the system.  One is
the generalized Abrikosov factor  
\be  
\beta_A= N_z L_x L_y \frac{\sum_{n} \int d^2{\bf r} | \psi_{n}({\bf 
r})|^4}{\left[ \sum_n 
\int d^2{\bf r} | \psi({\bf r})|^2\right]^2}\label{eq:beta}. 
\ee 
When the vortices form a triangular lattice, $\beta_A$ reaches 
its minimum value of $\beta_\triangle$, but exceeds
this value for other vortex configurations.  We have also computed 
the
spatial average of $|\psi_n({\bf r})|^2$, defined as
\be 
r_{ab}= \frac{1}{N_z L_x L_y} \sum_n \int d^2{{\bf r}} |\psi_n({\bf
r})|^2, \label{eq:delta} 
\ee 
which at low temperatures reaches the  mean field value 
$r_{ab}^{\mathrm{MF}}= |\alpha_H|/(\beta_\triangle \beta)$.
Both $\beta_A$ and $r_{ab}$ vary smoothly with temperature and thus do 
not show any special behavior at the flux lattice melting temperature 
$T_M(B)$.  We have therefore also examined three other equilibrium 
quantities which more clearly show signals of this transition: the 
isothermal shear modulus $\mu(T)$ of the flux lattice; a  
quantity we denote 
${\cal C}(T)$, which measures the degree of coherence between  
vortices in 
adjacent layers; and the zz-component of 
the helicity modulus tensor, $\Upsilon_{{\mathrm{cc}}}(T)$, which measures 
stiffness against a long-wavelength twist in the  phase 
of the order parameter.   
 
The shear modulus $\mu(T)$ is defined \cite{sasik2}  by
\be 
\mu(T)= \frac{1}{N_z L_x L_y}\left(\frac{\partial^2 
{F}}{\partial\theta^2}\right)_{T,\theta=0}, \label{eq:shear} 
\ee 
where $\theta$ is the shear angle. The free energy $F$ can be obtained
from Eq.\ (\ref{eq:ham1}), using
$F=-k_B T \ln Z$ where $Z=\mathrm{Tr}\, e^{-{\cal F}/k_B T}$ and
the trace is taken over the classical variables $c_{n,k}$ and
$c_{n,k}^*$.  An explicit $2D$ form for $\mu(T)$  in the LLL
approximation has been given in Ref.\ \onlinecite{sasik2}, where  
it has been found that  $\mu(T)$ reaches its mean field value
$\mu^{\mathrm{MF}}(T)=0.354 N_\phi k_B T g^{2}(T)$ at low $T$ and  it vanishes everywhere in the
liquid phase.  If the transition between the vortex solid and vortex
liquid state is first-order, then $\mu(T)$ will, in the thermodynamic
limit, jump {\em discontinuously} from a finite value to zero at
$T_M(B)$.  However, such a jump does {\em not} prove that the melting
transition is first-order, since certain continuous melting
transitions in two dimensions also have a
jump in $\mu(T)$ at melting.\cite{nelson}
However, other independent calculations give evidence 
that the melting transition is first-order within the 
LLL approximation in three dimensions (e.g., by exhibiting
a finite latent heat).
 
${\cal C}(T)$ is defined by 
\be 
{\cal C}(T) =\frac{ \sum_n \int d^2{\bf r} |\psi_{n+1}({\bf r}) 
-\psi_{n}({\bf r})|^2} 
{2\sum_n \int d^2{\bf r} |\psi_{n}({\bf r})|^2}. \label{eq:Gamma} 
\ee 
At low $T$, where the vortex system forms a flux lattice with 
flux lines all parallel to the c-axis,  
$\psi_{n+1}({\bf r}) = \psi_n({\bf r})$ and hence ${\cal C}=0$.  
By contrast, deep in the liquid phase, the phases of 
$\psi_n({\bf r})$ and $\psi_{n+1}({\bf r})$ are uncorrelated, 
and ${\cal C}$ approaches unity. To calculate ${\cal C}(T)$ and other ratios of
spatial averages, we evaluate the ratios at fixed time, and
then average over a period of time as described below.
 
Finally, the helicity modulus component $\Upsilon_{{\mathrm{cc}}}(T)$  
is defined by the relation\cite{fisher} 
\begin{equation} 
\Upsilon_{{\mathrm{cc}}}(T) = \frac{1}{V} 
\left(\frac{\partial^2 F}{\partial a_z^2}\right)_{T,V;a = 0}. 
\label{eq:upsilon} 
\end{equation} 
Here $a_z$ is an additional uniform vector potential applied in 
the z-direction (besides that which is needed to produce the 
magnetic field), and V is the system volume.  A further discussion 
of the meaning of $\Upsilon_{{\mathrm{cc}}}$ is to be found in Ref.\ 
\onlinecite{sasik1a}.  In the mean field approximation, 
$\Upsilon_{{\mathrm{cc}}}(T)$ is approximated by  
$\Upsilon_{{\mathrm{cc}}}^{\mathrm{MF}}(T) = 2Jd d_0 r_{ab}^{{\mathrm{MF}}}/\Phi_0^2$, 
where $r_{ab}^{{\mathrm{MF}}} = |\alpha_H|/(\beta_{\Delta}\beta)$ is the 
mean field value of the quantity $r_{ab}$ defined in Eq.\ 
(\ref{eq:delta}).  $\Upsilon_{{\mathrm{cc}}}$ is shown in Ref.\ \onlinecite{sasik1a} 
to drop discontinuously to zero at $T_M(B)$. 
 
\subsubsection{Dynamical Quantities} 
 
The wave-number- and frequency-dependent conductivity of 
the vortex system can be computed using the Kubo formula.
If the frequencies $\omega$ satisfy the 
condition $\hbar\omega \ll k_BT$, one may use the Kubo formula 
in the classical limit:\cite{kubo}
\bea
\sigma_{\mu\nu}({\bf q},\omega) &=& 
 \frac{1}{k_B T V} \int dt \int d^3{\bf x}\,  
d^3{\bf x'} \nn \\ &\,&e^{i {\bf q}\cdot ({\bf{x}}-{\bf {x'}})- i \omega t}
\la j_\mu({\bf x}, t) j_\nu({\bf x'},0) \ra. \label{eq:kubo} 
\eea
Here $j_\mu({\bf x}, t)$ is the 
$\mu${th} component of the current density, and  
$\sigma_{\mu\nu}({\bf q},\omega)$ is the $\mu \nu^{th}$ component 
of the complex conductivity tensor for a wave number ${\bf q}$ 
and frequency $\omega$, and  
$\la \cdots \ra $ denotes an average over the thermal noise 
distribution.

In the present work, we have considered only the 
conductivity component $\sigma_{c}$, which requires only 
the c-axis current density.  Within the Lawrence-Doniach model,  
this current density, for $z$ in the region between the $n${th} and 
$(n+1)${th} layer, is 
\be 
{j_{z}^{(n)} }({\bf r}, t)= \frac{\hb\,q }{m_c\,d}{\mathrm Im}\left[\psi^{*}_{n+1}({\bf 
r})\psi_{n}({\bf r})  \right]. \label{eq:current}  
\ee  
If we expand $\psi(r)$ using the representation (\ref{eq:LLL}), 
we find that 
$J_z(t)\equiv \frac{1}{N_z}\,\sum_{n} \int d^2{{\bf r}} j_{z}^{(n)}({\bf r},t)$ is  
\be 
J_z(t)=\frac{\hbar\,q\, L_x \ell \sqrt{\pi}}{m_c\,d}\, 
|\psi_{0}|^{2} {\mathcal{J}}(t), \label{eq:currentZ} 
\ee 
where  
\be 
{\mathcal{J}}(t)= \frac{1}{N_z}\,{\mathrm {Im }}\sum_{k,n} 
c^{*}_{n+1,k} c^{}_{n,k}.  
\label{eq:joft}
\ee  
Note that this current density includes only the Josephson 
currents between the layers, and not any additional normal currents 
which may be flowing in parallel.  
 
The corresponding real fluctuation conductivity $\sigma_{c,1}(\omega) 
\equiv {\mathrm{Re}}\left[\sigma_{cc}({\bf q}=0, \omega)\right]$   
follows from  the Kubo formula (\ref{eq:kubo}): 
\be 
\sigma_{c,1}({\omega})= \frac{d_{0}^{2} N_{z}^{2}}{k_B T V }
\int_{0}^{\infty} dt 
\cos(\omega t)\,\la J_z(t) J_z(0) \ra. 
\label{eq:kuboc} 
\ee  
Upon using Eq.\ (\ref{eq:currentZ}), it takes the form 
\be 
\frac{\sigma_{c,1}({\omega'})}{\sigma_0}= \frac{N_z n_x G^2(T)}{n_y}  
\int_{0}^\infty d\tau 
\cos(\omega' \tau)\,\la {\mathcal{J}}(\tau) {\mathcal{J}}(0) \ra, 
\label{eq:conductivity} 
\ee 
where $\sigma_0= q^2 \,t_0 |\psi_0|^2/ m_c$, $\omega'=\omega\,
\,t_0$, and
\be 
G^2(T)= \frac{2\, d_0\,\eta g^2(B,T)}{3^{1/4} d}.
\ee  
 
Besides the frequency-dependent conductivity, it is sometimes of 
interest to compute the {\em integrated fluctuation conductivity},  
$\gamma_2$, defined by\cite{ebner} 
\be 
\gamma_2= \frac{1}{\pi} \int_{0}^{\infty} d\omega' \,\sigma_{c,1}({\omega'}).  
\label{eq:gamma2} 
\ee 
With the use of Eq.\ (\ref{eq:conductivity}),  
$\gamma_2$ can be simplified to 
\be 
\gamma_2= \frac{G^2(T) \sigma_0 n_x}{n_y N_z}   
\left \la |{\mathcal{J}}(0)|^2 \right \ra \label{eq:gamma2a}, 
\ee   
where ${\mathcal{J}}(0)$ is given by Eq.\ (\ref{eq:joft}).

\begin{figure}[tb] 
\includegraphics[width=9.2cm]{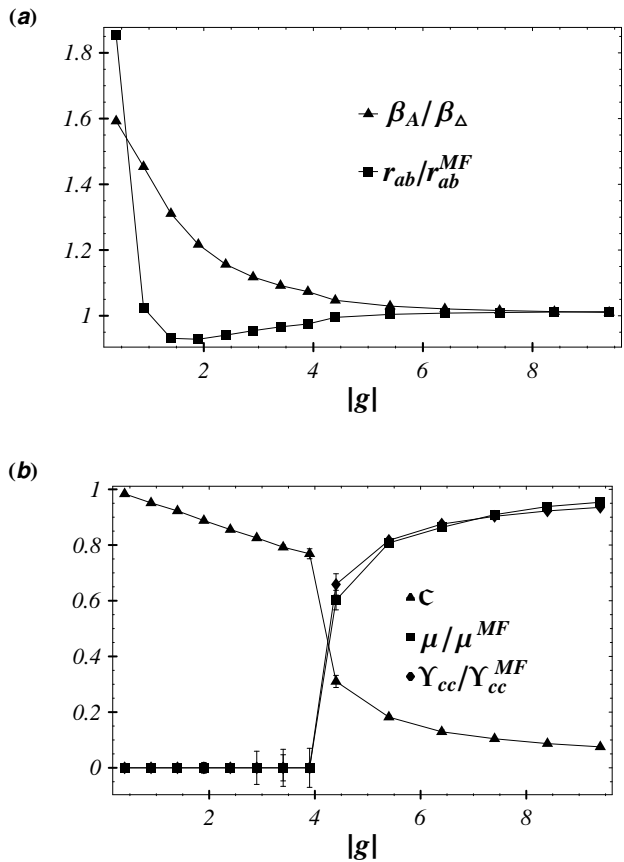} 
\caption{
\label{fig.mc}
(a) Ratio of the mean-square gap $r_{{\mathrm{ab}}}$ [Eq.\
(\ref{eq:delta})] to its mean-field value $r_{\mathrm{ab}}^{MF}$, and
the ratio of the generalized Abrikosov factor $\beta_A$ [Eq.\
(\ref{eq:beta})] to its value $\beta_\Delta$ in a triangular lattice,
plotted as a function of $|g|$ [Eq.\ (\ref{eq:g2})].  (b) The
calculated ratios of the shear modulus $\mu$ and the c-axis helicity
modulus $\Upsilon_{cc}$ to their mean-field values $\mu_{MF}$ and
$\Upsilon_{cc}^{MF}$, as obtained from Langevin dynamical simulations,
plotted as a function of $|g|$.  Also shown is ${\cal C}(T)$ [Eq.\
(\ref{eq:Gamma})].  The lattice used contains $N_\phi = 6 \times 6$
vortices in the ab-plane and $N_z = 12$ layers in the c-direction,
with periodic boundary conditions; the interlayer coupling is chosen
so that $\eta |g|=0.02$. The error bars in this and later figures
represent the standard deviations of results from about five Langevin
dynamical simulations run for equal lengths of time.}
\end{figure}

\section{Results}  
 
We have solved the Langevin equations (\ref{eq:langevin}) numerically
using a second order Runge-Kutta algorithm.  In this algorithm,
$\mathcal{F}$ is correct through ${\cal O}(\epsilon^2)$ in the time
step $\epsilon$. \cite{wilson} In most of our simulations, we used
$\epsilon=0.15\,t_0$; a smaller time step of $\epsilon=0.05 \,t_0$
was found to give similar results but to require more computer time.
The real and imaginary parts of the noise term $\xi_{n,k}'(\tau)$ in
Eq. (\ref{eq:langevin}) are chosen from Gaussian distributions with a
mean zero and a variance $\sigma^2/\epsilon$ where $ \sigma^2= 2/[n_x
g^2(B,T)]$.  This choice insures that these terms have mean and
variance which satisfy Eqs.\ (12) and (13).
 
In most cases, we have started our simulations from the
low-temperature Abrikosov phase, then gradually increased $T$, taking
the initial state for a higher $T$ as the equilibrium state for the
previous slightly lower $T$.  We have verified that our results
exhibit only a little  hysteresis - that is, we obtain the same equilibrium
and nearly the same dynamical results, whether $T_M(B)$ is approached
from below or from above. We have found that our choice of initial
state generally has little effect on dynamics, providing we ``anneal''
our sample for a long enough time as described in the next paragraph.
We have confirmed this lack of effect by obtaining similar results for
various calculated dynamical quantities whether we begin by choosing
an Abrikosov or a liquid-like initial state.

For each temperature considered, we have allowed the system to
equilibrate for a period ranging from $10^{3}$ to $4 \times 10^{6}$
time steps, before starting to compute averages, the larger number
corresponding to temperatures close to $T_M$.  
We then run the dynamics for an additional $3 \times 10^4$ to 
$10^6$ time steps at this temperature, and use
these results to compute the averages. 

To calculate the quantity of interest, we include in the 
averages only the results obtained in every $N_0$ time steps,  
where $N_0$ is chosen as explained below.  If this
procedure is used, then, according to Ref.\ \onlinecite{critical}, 
the consecutive values included in the average become nearly 
statistically independent.  We choose $N_0$ using
a criterion involving the so-called self-correlator.  This
self-correlator is defined by the relation $C_{x}(k)=( \langle x_{i+k}
x_i\rangle - \langle x_i \rangle^2)/( \langle x_i^2\rangle- \langle
x_i\rangle^2)$, where $x_i$ is the physical quantity of interest at
the $i$th time step, and $\langle...\rangle$ is a time average.  With
this definition, $C_x(0) = 1$; also, $C_x(k)$ decreases as $k$
increases.  The optimum choice of $k$ to be used in the simulations
(i.e., the optimum number of steps between those included in the
averages) is that which makes $C_x(k)$ as small as possible, typically
less than $0.05$.  For our simulations, we find that this optimum
value of $k\equiv N_0$ is typically between $20$ and $50$.  In general, we find
that collective properties such as conductivity require much longer
runs than single particle properties such as $r_{ab}$; the optimum
ratio of collective to single-particle running times itself depends on
the system size.

\begin{figure*}[tb] 
\includegraphics[width=12.2cm]{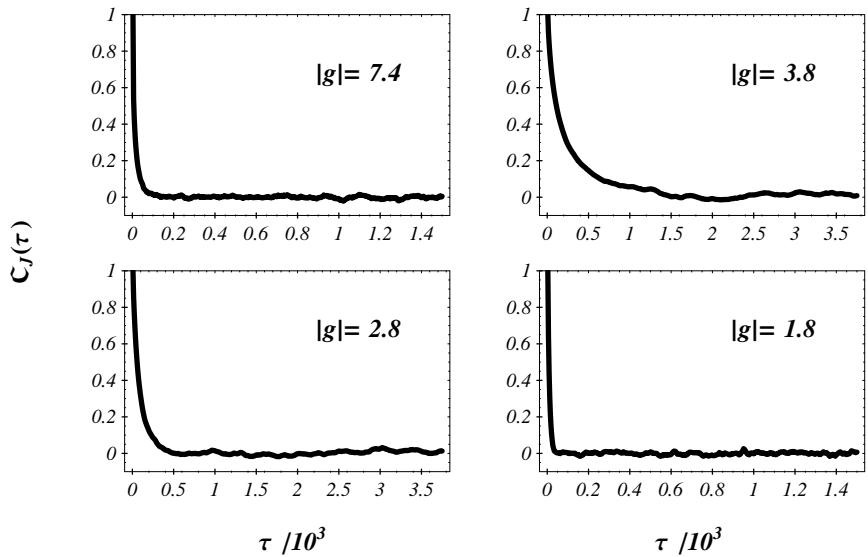} 
\caption{ 
\label{fig.corr}
Normalized correlation function ${\mathcal{C}}_J (\tau) = \la
{\mathcal{J}}(\tau) {\mathcal{J}}(0) \ra / \la {\mathcal{J}}(0)^2 \ra$
plotted as a function of $\tau$ for several values of $g$ as indicated
in the Figure. The freezing transition occurs at $|g| \approx 4$ for this
lattice size. Note the expanded horizontal scale. 
The lattice used here
has $16$ vortices in each plane, and $16$ planes ($16 \times 16$ lattice), 
and we choose $\eta |g|= 0.05$. }
\end{figure*}

As a test of our numerical algorithm, we have computed several
equilibrium properties of the system which have been previously
evaluated using Monte Carlo methods. \cite{sasik1a,hu} Our results are
shown in Fig.~\ref{fig.mc}.  For these calculations, the lattice used
contained $6 \times 6$ vortices in the ab-plane and $12$ layers in the
c-direction. The coupling between the layers is chosen as $\eta
|g|=0.02$.  We emphasize that our calculations are intended to probe a
range of physically reasonable parameters, rather than to describe any
specific superconductor.  In Ref.~\onlinecite{hu}, $\eta|g|$ was
estimated to vary from $0.0075$ (BSCCO) to $0.30$ (YBCO) in typical
cuprate superconductors at temperatures and magnetic fields where the
LLL approximation is likely to be valid; our choice falls well within
this range.
 
Figure~\ref{fig.mc}(a) displays the generalized Abrikosov factor
$\beta_A/\beta_{\triangle}$ [Eq. (\ref{eq:beta})] and the quantity
$r_{ab}$ [Eq.\ (\ref{eq:delta})] as a function of $g$ [Eq.\
(\ref{eq:g2})].  These results are similar to previous results
obtained using the Monte Carlo method.\cite{hu} In
Fig.~\ref{fig.mc}(b) we show the calculated $\mu/\mu^{{\mathrm{MF}}}$
[Eq.\ (\ref{eq:shear})] versus $|g|$.  The sharp drop in
$\mu/\mu^{{\mathrm{MF}}}$ near $|g| \approx 4 $ is clearly visible.
Although our sample sizes are quite small,
the calculated equilibrium quantities still show the expected behavior
of larger samples, though somewhat broadened by the substantial finite
size effects.

Also shown in Fig.~\ref{fig.mc}(b) is the interlayer coupling strength
parameter ${\cal C}(T)$ [Eq.\ (\ref{eq:Gamma})] and the helicity
modulus component $\Upsilon_{{\mathrm{cc}}}(T)$, both plotted as
functions of $|g|$.  Both of these quantities (which are sensitive to
z-axis coherence) show a drop near $|g| \approx 4$, but ${\cal C}(T)$
is expected to vary smoothly through this region of $|g|$, while
$\Upsilon_{{\mathrm{cc}}}(T)$ is expected to drop discontinuously to
zero in the thermodynamic limit; some evidence of this distinction can
be seen in the Figure.  The simultaneous drop in
$\mu/\mu^{{\mathrm{MF}}}$ and $\Upsilon_{{\mathrm{cc}}}(T)$ near $T_M$
suggests that there is simultaneous flux lattice melting in the
ab-plane and interlayer decoupling in the c-direction, near $T = T_M$,
consistent with previous calculations in clean
systems. \cite{sasik1,sasik1a,hu,sasik2}

Next, we turn to the dynamics of the system. We have calculated $\left
\la J_z(t/\,t_0) J_z(0) \right \ra $ as a function of various system
parameters.  In Fig.~\ref{fig.corr}, we plot this correlation function
as a function of $t/\,t_0$ for several values of $|g|$ both above and
below the expected melting point, denoted $|g_M|$. ($|g_M| \approx 4$
for lattices of this size and our choice of
parameters\cite{sasik1,hu}).  Clearly, the decay rate slows
considerably as melting is approached from higher temperatures, 
i.e., from smaller values of
$|g|$. However, the decay rate is rapid and only weakly dependent on
$|g|$ in the vortex lattice phase, $|g| > |g_M|$.

\begin{figure}[tb] 
\includegraphics[width=9.2cm]{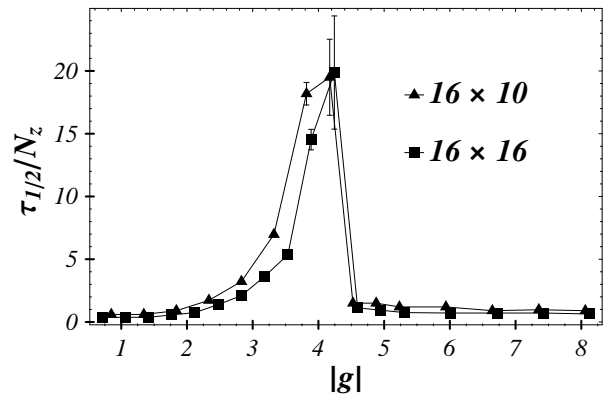} 
\caption{
\label{fig.hf}
Half-life $\tau_{1/2}$ characterizing the decay rate of $\langle
J_z(t/\,t_0)J_z(0)\rangle$ plotted as a function of $|g|$.  The full
lines simply connect the calculated points.  Error bars have the same
meaning as in Fig.~\ref{fig.mc}.  The sizes of the lattice used are
indicated in the Figure legend and the interlayer coupling is chosen
so that $\eta |g|=0.05$.}
\end{figure}

To make this $|g|$-dependence more apparent, we plot in
Fig.~\ref{fig.hf} the half-life $\tau_{1/2}$ of this correlation
function versus $|g|$ for two different system sizes, normalized by
$N_z\,t_0$.  $\tau_{1/2}$ is defined as the time at which $\left \la
J_z(t/\,t_0) J_z(0) \right \ra $ has fallen to half its $\tau = 0$
value.  Consistent with Fig.~\ref{fig.corr}, $\tau_{1/2}$ is
relatively small in both the solid and the liquid phases far from
$|g_M|$; it increases noticeably as $|g_M|$ is approached from the
liquid but not from the solid phase.  Despite this increase, we
believe that $\tau_{1/2}$ will not diverge at $|g_M|$, because $|g_M|$
corresponds to a first-order melting transition, with no dynamical
critical phenomena such as a diverging correlation time.

\begin{figure}[tb] 
\includegraphics[width=9.2cm]{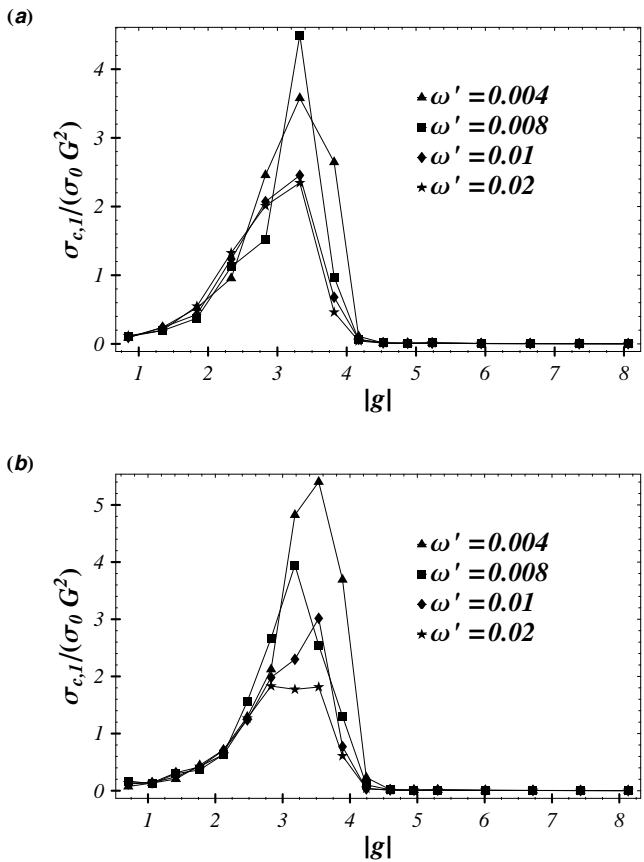} 
\caption{ 
\label{fig.cond}
(a) The real part of the fluctuation conductivity,
$\sigma_{c,1}(\omega^\prime)/[\sigma_0 G^2(T)]$, plotted versus $|g|$
for several values of $\omega^\prime$ for a 
$4 \times 4 \times 10 \equiv 16 \times 10$ lattice.
(b) Same as (a) except that the lattice size is 
$ 4 \times 4 \times 16$.  Note
the sharp increase in $\sigma_{c,1}(\omega^\prime)$ in the vortex
liquid phase near and below $|g_M|$ for small $\omega^\prime$. The
parameters are the same as in Fig.~\ref{fig.corr}.  For clarity,
we do not show error bars; they are comparable to those of
Fig.~3.}
\end{figure}

Figures~\ref{fig.cond}(a) and ~\ref{fig.cond}(b) show
$\sigma_{c,1}(\omega')$, as obtained from Eq.\ (\ref{eq:kuboc}) for
several values of $\omega^\prime$, and for two different system sizes.
The lines simply connect the calculated points.  The melting value
$|g_M| \approx 4$, as estimated from Fig.\ 1(b).  For $\omega^\prime =
0.004$ and $0.008$, $\sigma_{c,1}(\omega^\prime)$ increases strongly
as $|g_M|$ is approached from the liquid side.  There is a smaller
increase at higher $\omega^\prime$, because the transition primarily
affects fluctuations on a lower frequency scale.  In the solid phase,
there is little evidence of fluctuations in
$\sigma_{c,1}(\omega^\prime)$, which remains very small at nonzero
frequencies for all $|g| > |g_M|$ studied.  At fixed $|g|$ in the
liquid phase near $|g_M|$, we expect $\sigma_{c,1}(\omega^\prime)$ to
decrease monotonically with increasing $\omega^\prime$; we ascribe any
deviation from monotonic behavior in Fig.\ 4 is to  numerical
uncertainties.  Similarly,
although $\sigma_{c,1}(\omega^\prime)$ sometimes seems to peak at a
value of $|g|$ slightly smaller than $|g_M|$, we believe that this
behavior also lies within our numerical uncertainties. 

\begin{figure}[tb]
\includegraphics[width=9.2cm]{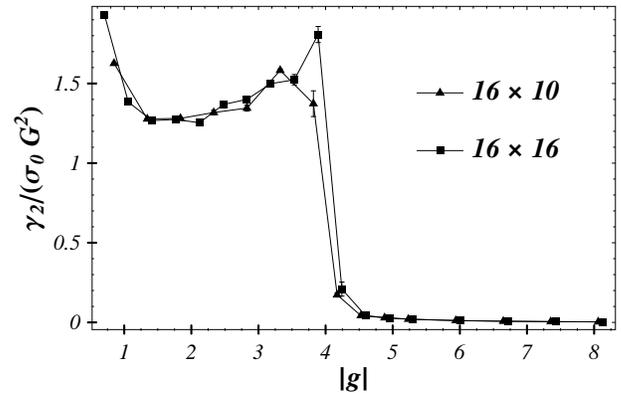} 
\caption{
\label{fig.isigma}
The integrated fluctuation conductivity $\gamma_2/[\sigma_0 G^2(T)]$
plotted as a function of $|g|$ for two lattice sizes as shown in the
Figure legend. The parameters are the same as in Fig.~\ref{fig.corr}}
\end{figure}

In Fig.~\ref{fig.isigma} we show the total integrated fluctuation
conductivity $\gamma_2$ [Eq.\ (\ref{eq:gamma2a})], in units of
$\sigma_0 G^2(T)$, plotted as a function of $|g|$ for two lattice
sizes.  We have computed $\gamma_2$ using the equilibrium expression
Eq.\ (\ref{eq:gamma2a}), which is equivalent to the frequency integral
of the quantity shown in Fig.~\ref{fig.cond}(a) or ~\ref{fig.cond}(b).
As $|g_M|$ is approached from the liquid side; $\gamma_2$ falls
sharply at $|g_M|$, to a small value in the vortex lattice phase.
This drops is expected to be discontinuous in the thermodynamic (large
size) limit.  As previously, the full lines simply connect the
calculated points. Note that $\gamma_2/[\sigma_0 G^2(T)]$  is
approximately $|g|$-independent in the liquid phase, because of the
way it is normalized ($G^2 \propto g^2$).

\section{Discussion}
 
The present results are consistent with the scenario of a first-order
melting transition at $T_M(B)$, where long-range vortex order in the
ab-planes, and in the c-direction, disappear simultaneously.  This
interpretation is supported by the behavior of the helicity modulus
$\Upsilon_{{\mathrm{cc}}}(T)$ and shear modulus $\mu(T)$, both of
which vanish at the same temperature $T$.  Although the present
calculations are limited to relatively small samples (with fewer than
$50$ vortex pancakes per plane), similar behavior has been observed in
Monte Carlo simulations for considerably larger
systems. \cite{sasik1,hu,sasik1a}

The behavior of {\em dynamical} properties, such as
$\sigma_{c,1}(\omega^\prime)$, is also consistent with first-order
melting.  For small $\omega^\prime$, $\sigma_{c,1}(\omega^\prime)$
shows a strong increase as $T$ approaches $T_M$ from above. This
behavior occurs because, in the solid phase, the vortex pancakes in
adjacent layers lie above one another; as a result, fluctuating
currents between the layers are small.  On the other hand, when the
lattice melts (low $|g|$ or high $T$), the vortex pancakes in adjacent
layers no longer lie directly above one another; hence, fluctuating
phase gradients between the layers increase the current fluctuations
and the fluctuation conductivity. In the liquid phase,
$\sigma_{c,1}(\omega^\prime)$ decreases with decreasing $|g|$, 
because $\tau_{1/2}$ is becoming smaller.  However,
$\sigma_{c,1}(\omega^\prime)$ once again appears to show a
discontinuity rather than a divergence at $|g_M|$, consistent with the
first-order nature of the transition.

Why does $\sigma_{c,1}(\omega^\prime)$ in the liquid state, shown in
Fig.~\ref{fig.cond} for several frequencies, decrease with increasing
temperature $T$ above $T_M$?  We believe this decrease occurs because,
as shown in Fig.~\ref{fig.hf}, $\tau_{1/2}$ decreases with increasing
$T$.  By contrast, the quantity $\gamma_2/(\sigma_0 G^2)$, shown in
Fig.~\ref{fig.isigma}, behaves differently from $\sigma_{c,1}$: it has
a discontinuity at $T_M$ but varies slowly in the liquid.  The lack of
any clear peak in $\gamma_2/(\sigma_0 G^2)$ near $T_M$ can be
understood from Eq.\ (\ref{eq:gamma2}), which shows that $\gamma_2$ is
independent of $\tau_{1/2}$, depending only on equal-time current
density fluctuations at $t=0$.  By contrast,
$\sigma_{c,1}(\omega^\prime)$ is sensitive to $\tau_{1/2}$ 
especially for small  $\omega^\prime$.

At this point, we briefly comment on a seemingly counterintuitive
feature of the dynamical results shown in Figs.~\ref{fig.cond},
namely, that $\sigma_{c,1}(\omega^\prime)$ is small in the vortex
lattice phase for nonzero $\omega^\prime$, even well below $T_M$.
Intuitively, one might expect, since this phase is superconducting,
with a finite helicity modulus in the $c$ direction, that
$\sigma_{c,1}(\omega^\prime)$ would be large in this regime.  However,
this behavior is actually physically reasonable; our picture of the
underlying physics is the following.  We believe that
$\sigma_{c,1}(\omega^\prime)$ corresponding to our model dynamics is
the sum of two parts: (i) the fluctuation conductivity shown in
Figs.~\ref{fig.cond}(a) and ~\ref{fig.cond}(b) (whose integral is
shown in Fig.~\ref{fig.isigma}; and (ii) a delta function at zero
frequency, corresponding to perfect conductivity.  The delta function
does not appear in Fig.~\ref{fig.cond}(a) or ~\ref{fig.cond}(b)
because those calculations are carried out at finite frequency, nor
does it appear in the integral shown in Fig.~\ref{fig.isigma}.  The
strength of this delta function is proportional to the helicity
modulus shown in Fig.~\ref{fig.mc}(b), which vanishes for $T > T_M$.
Thus, $\sigma_{c,1}(\omega^\prime)$ is small below $T_M$ simply
because the {\em fluctuation} contributions to the conductivity are
small in this temperature range; the system is still phase coherent in
the c-direction and still has a finite helicity modulus below $T_M$.
Although it may seem strange that $\sigma_{c,1}(\omega^\prime)$ is
small for $T < T_M$ and finite $\omega^\prime$, this behavior is not
unprecedented.  For example, in low-T$_c$ s-wave superconductors, the
existence of a finite gap below $T_c$ means that $\sigma_1(\omega) =
0$ for $T < T_c$ and for  $\hbar\omega$ smaller than twice the energy gap.
  
To our knowledge, no direct measurements of
$\sigma_{c,1}(\omega^\prime)$ have been carried out in the cuprate
superconductors in the high-field, clean-limit regime where our
calculations might be applicable.  We therefore comment briefly on an
entirely different experiment in which the reported behavior somewhat
resembles that shown in Fig.~\ref{fig.cond}.  This is a recent study
of the frequency-dependent conductivity of \bscco within the ab plane
at zero applied magnetic field. \cite{corson}  This experiment reports
a rather sharp peak near $T_c$ in the real part of the in-plane
conductivity at about 0.2 THz.  This peak is thought to be due to
fluctuations in the phase of the order parameter which are strongest
near $T_c$, and weaker both above and below $T_c$.  We believe that
similar fluctuations (probably in the amplitude of the order parameter
as well as the phase) are producing the increase in
$\sigma_{c,1}(\omega^\prime)$ in the present model near $T_M$.  These
fluctuations are, we believe, limited in size because the melting
transition is first-order rather than continuous, and they are
relatively small for $T < T_M$.

The present
calculations may be relevant to c-axis transport at {\em strong}
magnetic fields (where the LLL approximation is adequate) in a {\em
clean} high-T$_c$ superconductor (where a first-order vortex lattice
melting is expected), provided a Langevin dynamics is appropriate.  We
have tried to estimate the numerical value of our calculated
$\sigma_{c,1}(\omega^\prime)$ 
[Eq.\ (\ref{eq:conductivity})] for reasonable
experimental parameters.  
For $\omega^\prime \ll 1$ or $ \omega t_0  \ll 1$, 
$\sigma_{c,1}(\omega^\prime)$ has the
same order of magnitude as the quantity
\begin{equation}
\sigma_0\,G^2(T) = q^2\,t_0(2d_0/\pi\ell^2\hbar^2)k_BT 
(\eta g)^2 g^2.
\label{eq:sigg2}  
\end{equation}
All the quantities in this equation are easily determined except
$\,t_0$.  We attempt to estimate $\,t_0$ using an early paper by
Schmid,\cite{schmid} in which the time-dependent Ginzburg-Landau
equation is derived from the original BCS theory within the Gor'kov
approximation.  Schmid finds that $\,t_0 \approx h/[32\,k_B
T_{c0}(1-T/T_{c0})(1-B/H_{c2})]$, where $T_{c0}$ is the mean-field
superconducting transition temperature at $B = 0$.  Taking $T_{c0}
\sim 80$~K, $T \sim~60$~K we find $\,t_0 \sim 4~\times~10^{-14}$~sec
at a field of $5$~T.  Substituting this value into Eq.\
(\ref{eq:sigg2}), and using $\eta g = 0.05$, $g^2 \sim 20$, and $d_0
\sim 5$~\AA, we obtain $\sigma_0 G^2 \sim 5 \times 10^{11}$~esu, or
about 0.06~$\Omega^{-1}$~cm~$^{-1}$.  This conductivity is
considerably smaller than the apparent c-axis conductivity in the
vortex liquid state, even in a very anisotropic material such as
\bscco. \cite{fuchs} Thus, it might be difficult to observe the
fluctuation contribution in a clean anisotropic superconductor, unless
we have substantially underestimated $\,t_0$.  Such an underestimate
is possible, since the calculations of Schmid are based on a
microscopic theory which may not be directly applicable to the layered
high-T$_c$ materials.

Few experiments appear to have measured the c-axis resistivity at the
high fields where the LLL approximation would be most accurate.  Fuchs
{\it et al},\cite{fuchs} working at far lower fields ($\sim 25$-$200$
~Oe), have observed a simultaneous disappearance of resistivity in the
ab-plane and in the c-direction (indicating a single phase
transition), and an abrupt increase in the c-axis resistivity at a
temperature just above that transition.  However, their experiments
are done at such frequencies ($ \approx 72$~Hz) that the inductive
contribution is very dominant in the solid phase.
 
Finally, we briefly discuss the fact that our calculated hysteretic
effects are very small in the vicinity of the first-order melting
transition.  In a real experiment, one might expect some evidence of
superheating or supercooling.  This minimal amount of hysteresis may
be due to the long annealing time before we begin to calculate
thermodynamic averages.  Because of this long annealing, our system
can apparently attain its thermodynamic state of minimum free energy
{\em before} we start computing averages.

In summary, we have studied both the equilibrium and the dynamical
behavior of a layered superconductor in a strong magnetic fields by
solving the time-dependent Ginzburg-Landau equations, in the lowest
Landau level approximation.  The effects of fluctuations are
incorporated by means of a Langevin noise term.  The equilibrium
properties are found to exhibit behavior similar that found in
previous Monte Carlo results: \cite{sasik1,sasik1a,hu,sasik2} a
first-order melting transition of the vortex lattice, with a
simultaneous loss of in-plane and interplane order.  The dynamical
properties show a strong, but not divergent, increase in the c-axis
conductivity as $T_M$ is approached from above, with a corresponding
increase, but no divergence, in the half-life $\tau_{1/2}$ of the
c-axis current fluctuations.

\section{Acknowledgments} 
 
We gratefully acknowledge support through NSF grant DMR01-04987.



\begin{thebibliography}{3}
\bibitem{note0} Here we assumed that the lowest Landau level
eigenstates $\phi_k(x,y)$ form a complete set.
\end{thebibliography}
\end{document}